\documentclass[12pt]{article}
\usepackage{graphicx}

\begin{document}
\thispagestyle{empty}
\vskip 10mm
\begin{center}
{\Large\bf
Off-diagonal quark distribution functions of the pion within
an effective single instanton approximation.
}

\vskip 5mm
I.V. Anikin$^{ac}$, \, A. E. Dorokhov$^{ab}$,
\, A.E. Maximov$^a$,
\, L. Tomio$^b$, \, V. Vento$^c$

\vskip 5mm

{\small {\it $^{a}$Bogoliubov Laboratory of Theoretical Physics, Joint
Institute for Nuclear Research,  Dubna, Russia} }\\
\centerline{\tt anikin@thsun1.jinr.ru }
\centerline{\tt dorokhov@thsun1.jinr.ru }
\centerline{\tt maximov@thsun1.jinr.ru }
\vskip 0.3cm

{\small {\it $^{b}$ Instituto de F\'\i sica Te\'orica, Universidade Estadual
Paulista,\\ S\~{a}o Paulo, Brasil} } \\

\centerline{\tt tomio@ift.unesp.br }
\vskip 0.3cm

{\small {\it $^{c}$ Departamento de Fisica Teorica and Instituto de
F\'{\i}sica Corpuscular,\\
Universidad de Valencia - Consejo Superior de Investigaciones
Cient\'{\i}ficas,\\ Valencia,   Spain} }\\
\centerline{\tt vicente.vento@uv.es}

\end{center}
\vskip 5mm

\begin{center}
\begin{minipage}{150mm}
\centerline{\bf Abstract}
\vskip 5mm

We develop a relativistic quark model for pion structure, which
incorporates the non-trivial structure of the vacuum of Quantum
Chromodynamics as modelled by instantons. Pions are boundstates of
quarks and the strong quark-pion vertex is determined from an
instanton induced effective lagrangian. The interaction of the
constituents of the pion with the external  electromagnetic field is
introduced  in gauge invariant form. The parameters of the model,
i.e., effective instanton radius and  constituent quark masses, are
obtained from the vacuum expectation  values of the lowest
dimensional quark and gluon operators and  the  low-energy
observables of the pion. We apply the formalism to the  calculation
of the pion form factor by means of the isovector nonforward
parton  distributions
and find agreement with the experimental data.

\end{minipage}
\end{center}

\vskip 1cm

\leftline{Pacs: 11.15.Kc, 12.38.Lg,12.39.Ki,14.65.Bt}

\leftline{Keywords: vacuum, instanton, quark, meson, parton.}

\newpage

\section{ Introduction}

\indent The quark and gluon distribution functions play an important
role in the exploration of hadron structure. Perturbative quantum
chromodynamics (pQCD) allows one to obtain the evolution of the
distribution functions with $Q^2$ using operator product expansion
and renormalization group methods from some starting input, which
cannot be calculated from the first principles. We have been
therefore motivated to develop an effective approach which, based on
fundamental principles of QCD, serves to estimate the observables
associated with the dynamics of large distances.

Recently much attention has been paid to the off-diagonal quark (and
gluon) distribution functions parametrizing asymmetrical hadron
matrix elements of the light-cone quark-gluon operators. The
off-diagonal distribution functions are a generalization of the
conventional distribution functions, which serve as a link between
the hadronic structure functions and their form factors.

In here, we calculate the isovector off-diagonal leading-twist quark
distribution functions for the pion within a novel approach, which is
an extension of the well-known instanton model
\cite{Dia1}-\cite{Shu2} incorporating bosonization.
An effective
quark-hadron lagrangian of nonlocal type, where the nonlocality is
generated by instantons, underlies our approach. We implement gauge
invariance with respect to external electromagnetic field in the
effective lagrangian and the Green functions by introducing
explicitly the Schwinger phase factor in the definition of quark
fields. The parameters of the instanton vacuum, the instanton radius
and the constituent quark mass, are related to vacuum expectation
values of the lowest dimension quark-gluon operators. The coupling
constants are calculated from the compositeness condition which is
formulated in our case as the vanishing of the renormalization
constant of the meson-field. The quark distribution functions
obtained correspond to a hadronic scale (point of normalization)
given by the typical relative quark momenta that flows through the
instanton vertex and is of order $\mu_0\sim p\sim\rho_c^{-1}\approx
0.5\div 1\,\,GeV$.
We ought to stress that at some point in our development we restrict
the calculation to the
isovector part of distribution which is the only one required by
the scrutinized observable \footnote{A more complete analysis in a
different model can be found in ref.\cite{PW}.}.

We proceed in the next section to describe the nonlocal four-quark
model induced by instanton exchange in the effective single instanton
approximation.

Section 3 is devoted to fix the parameters of the model from the
experimental values of the vacuum condensates and the properties of
the pion. In section 4 we calculate the corresponding nonforward parton
distributions associated with deeply virtual Compton scattering and
determine the pion form factor. In an appendix we explicitly show the
contributions from the different diagrams.

\section{ Model}

\indent The investigation of the ground state properties of QCD has
important consequences for the description of the hadronic processes
within the theory. At present, there exists at least three main
approaches for studying the structure of the vacuum: i) the
quasi-classical approximation; ii) a formalism based on Wilson's
operator product expansion ;  iii) lattice QCD. As it is well-known,
the main problem of the quasi-classical method is that, for the
$3+1$- dimensional Yang-Mills theory, the instanton gas approximation
is not able to provide the correct vacuum-vacuum amplitudes in the
background of large QCD-vacuum fluctuations. In a scale-invariant
theory, like QCD, the vacuum fluctuations can have large sizes
leading to an infinite instanton--anti-instanton medium density, this
is the so called infrared divergence problem, which hinders the
solution.

At the present, one way for stabilizing the pseudoparticle medium is
the instanton liquid model ( see, for example, ~\cite{Dia1,Dia3}). In
the framework of instanton liquid model, the medium of $I-\bar{I}$ is
quite rarefied, i. e., the ratio of the average size to the
average distance between pseudoparticles being approximately equal to
$1/3$. Besides, in the limit $(N_{c}\rightarrow \infty )$ the size of
all (anti-)instantons is taken to be equal to some average size
$\bar{\rho}$. The instanton model with fixed average size of
instantons and instanton density is the effective single instanton
approximation of the QCD vacuum .

Within the single instanton approximation the chirally invariant
nonlocal four-fermion action has the following form \cite{Sch&Sh}
\begin{eqnarray}
\label{Sinst}
&&S_{4q}=\frac{1}{2}G_{I}\int d^{4}X_0\int \prod_{n}d^{4} x_{n}
K_{I}(x_{1},x_{2},x_{3},x_{4}) \nonumber\\
&&\left\{ \sum_{i}\left[ \overline{Q}_{R}(X_0-x_{1},X_0) \Gamma_{i}
Q_{L}(X_0,X_0+x_{3})\right] \left[ \overline{Q}_{R} (X_0-x_{2},X_0)
\Gamma_{i}Q_{L}(X,X+x_{4})\right] \right. \nonumber\\
&&\left. +(R\leftrightarrow L)\right\}
\end{eqnarray}
where $X_0$ is the position of the center of the instanton, and the
matrix combinations $\Gamma_i\times\Gamma_i$ are
\begin{eqnarray}
(\tau ^{a}\otimes \tau ^{a}),\qquad \frac{1}{2\left( 2N_{c}-1\right)
}(\tau ^{a}\sigma _{\mu \nu }\otimes \tau ^{a}\sigma _{\mu \nu }),
\end{eqnarray}
$\tau ^{a}=(1,i\vec{\tau})$ are the matrices characterizing
flavor space and  $N_{c}=3$ is the number of colors. To ensure the gauge
invariance of the nonlocal action (\ref{Sinst}) with respect to
an external electromagnetic $A_\mu(z)$ field, we define, following
\cite{Man,Tern91}, the chiral quark fields with the Schwinger phase
factor
\begin{eqnarray}
&&Q_{R(L)}(x,y)=\frac{1\pm \gamma_{5}}{2}Q(x,y) \nonumber\\
&&Q(x,y)={\rm exp}\left( -ieQ\int\limits_{x}^{y}dz_{\mu } A^{\mu}(z)
\right)q(y) \nonumber\\
&&Q={\rm diag}\left( \frac{2}{3},-\frac{1}{3},-\frac{1}{3} \right).
\end{eqnarray}
In the local limit the action (\ref{Sinst}) describes
quarks interacting through the 't Hooft vertex~\cite{tH} and it is
invariant under global axial ($q(x)\rightarrow \exp {(i\gamma
_{5}\tau \cdot \theta )}q(x)$) and vector ($q(x)\rightarrow \exp
{(i\tau \cdot \theta )}q(x)$) transformations, and it anomalously
violates the $U_{A}(1)$ symmetry ($ q(x)\rightarrow \exp {(i\gamma
_{5}\theta )}q(x)$). In what follows we neglect the current quark
mass and restrict ourselves only to the nonstrange quark sector. In
(\ref{Sinst}), the kernel of the four-quark interaction
$K_{I}(x_{1},x_{2},x_{3},x_{4})$ has the form
\begin{eqnarray}
\label{Kern Separable}
K_{I}(x_{1},x_{2},x_{3},x_{4})=f(x_{1})f(x_{2})f(x_{3})f(x_{4}),
\end{eqnarray}
where $f(x_{i})$ are related to the profile function
of the quark zero modes (see below). Morever, this kernel characterizes
the range of the nonlocality induced by quark-(anti-)quark instanton
interaction, which is determined by the average size of instantons.
The latter represents a natural cutoff parameter of the
effective low energy theory.

The $4$-fermion nonlocal action (\ref{Sinst}) is linearized
by introducing some auxiliary field interpreted as pion field.
Since this bosonization procedure has been elucidated in the literature
(see, for details, \cite{Vol}), let us immediately turn to
the linearized effective quark-pion
interaction lagrangian in
gauge-invariant form which is given by
\footnote{In a similar manner we can write the interaction lagrangian
for other mesons.},
\begin{eqnarray}
\label{Lag}
{\cal L}_{\pi\bar q q}^{{\rm
int}}(x)=g_{\pi\bar{q} q }\int d\xi _{1}d\xi _{2}f(\xi _{1})f(\xi _{2})
\overline{Q}(x+\xi _{1},x) i\gamma_{5}\tau\cdot\pi(x) Q(x,x-\xi _{2}).
\end{eqnarray}
The physical pion-quark coupling constant $g_{\pi\bar q q}$ is
determined from
a so-called coupling constant condition \cite{CoCo}
\begin{eqnarray}  \label{CopCon1}
Z_{\pi}=1+g^2_{\pi\bar\psi\psi}\Pi^\prime(m_{\pi}^2)=0,
\end{eqnarray}
where $Z_{\pi}$ is the renormalization constant of pion field. The physical
consequence of the condition $Z_{\pi}=0$ is that the pion field is always
found in a dressed state. Besides, the coupling constant condition is a
strong restriction because it implements the dynamics responsible for
the formation of the bound state.

Summarizing, lagrangian (\ref{Lag}) and the coupling constant condition
(\ref{CopCon1}) characterize our approach.

\section{Model parameters}

We briefly discuss the model parameters (for details, see ref.~\cite{DorL}).
Due to the effect of spontaneous breaking of the chiral symmetry, the
momentum dependent quark mass $M_{q}(k)$ is dynamically generated. It obeys
the well-known gap equation \cite{Dia1}
\footnote{
Here and below, all Feynman diagrams are calculated in Euclidean space
$(k^{2}=-k_{E}^{2})$ where the instanton induced form factor is well
defined, and
where it rapidly decreases so that no ultraviolet divergences arise. At the
very end we simply rotate back to Minkowski space. One can verify that the
numerical dependence of the results on the pion mass and the current quark
mass is negligible and can be safely ignored in the following
considerations.}, given by
\begin{equation}
\int \frac{d^{4}k}{(2\pi )^{4}}\frac{f^2(k)M_{q}(k)}{k^{2}+M_{q}^{2}(k)} =
\frac{n_{c}}{4N_{c}M_q},  \label{15}
\end{equation}
where $M_{q}(k)$, the momentum dependent quark mass, is defined by
the dressed quark propagator:
\begin{equation}
S_{F}^{-1}(p)=\rlap/p-M_{q}\tilde{Q}(p),  \label{q prop}
\end{equation}
and, $n_{c}=M_{q}^{2}/G_I$
is the density of the instanton medium ($G_I$ is the four-fermion interaction
constant in (\ref{Sinst}) and $M_q=M_q(0)$).
The solution of eq.(\ref{15}) has the
form
\begin{equation}
M_{q}(k)=M_{q}\tilde{Q} (k),  \label{16}
\end{equation}
where $\tilde{Q} (k)=f^2(k)$ is the nonperturbative part of the quark
propagator which in the axial gauge is given by
\begin{eqnarray}
\tilde{Q}(p) &=&\frac{1}{2(2\pi )^{2}}\frac{p^{2}}{\rho _{c}^{2}}\int
d^{4}x\ exp{(-ip\cdot x)}Q(x^{2}),\;\;\;\tilde{Q}(0)=1;\;\;\;(p=|p|)\;\;\;,
\label{Qin}
\end{eqnarray}
\noindent and
\begin{eqnarray}
Q(x^{2}) &=&\langle :\bar{q}(0)q(x):\rangle /\langle :\bar{q}
(0)q(0):\rangle ,\;\;\;\;Q(0)=1\;.  \label{Qx2}
\end{eqnarray}
In the axial gauge Eq. (\ref{Qx2}) coincides with the gauge-invariant
correlator in which the quark fields are connected by
the Schwinger phase factor.
In practical calculation we shall use the approximation for
(\ref{Qin}, \ref{Qx2}) as in ref.\cite{DorL}
\begin{eqnarray}  \label{9}
f(k)\approx 2\pi\rho_c\Biggl( 2.25\exp\left\{ -\rho_c | k | \right\} -
1.25\exp\left\{ -3\rho_c | k | \right\}\Biggr).
\end{eqnarray}

Given the value of the dynamical mass, one can obtain the value of the quark
condensate,
\begin{equation}
\langle \bar{q}q\rangle =\lim_{y\rightarrow x}{\rm tr}S_{F}(x-y)=-4N_{c}\int
\frac{d^{4}k}{(2\pi )^{4}}\frac{M_{q}(k)}{k^{2}+M_{q}^{2}(k)},  \label{17}
\end{equation}
and the average quark virtuality in the vacuum \cite{MihRad92a}-\cite
{MihRad92b},
\begin{equation}
\lambda _{q}^{2}\equiv \frac{\langle :\bar{q}D^{2}q:\rangle }
{\langle :\bar{q}q:\rangle }= -\frac{4N_{c}}{\langle
\bar{q}q\rangle }\int \frac{d^{4}k}
{(2\pi )^{4}}k^{2}\frac{M_{q}(k)}{k^{2}+M_{q}^{2}(k)}.  \label{18}
\end{equation}
The ratio $\eta =M_{q}^{2}(\lambda _{q}^{2})/\lambda _{q}^{2}$ characterizes
the diluteness of the instanton liquid vacuum.

Here, we have to emphasize that in the axial gauge the gauge invariant
quantity $M_q(k)$ is expressed in terms of the gauge invariant nonlocal
codensate (\ref{Qx2}). Thus, the quark condensate and
the average quark virtuality in the vacuum  are defined in
the gauge invariant manner.

For the moment, it is instructive to consider equations (\ref{15}) - (\ref
{18}) neglecting $M_{q}^{2}(k)$, compared to $k^{2}$, in the denominator
of the integrands. This approximation is justified in the dilute liquid
regime where $\langle k^{2}\rangle \sim \lambda _{q}^{2}>>M_{q}^{2}(\lambda
_{q}^{2})$. Using the explicit form in the momentum representation of the
zero mode in the axial gauge \cite{DorL}  from (\ref{17}), (\ref{18}) we
have  \cite{Dor}
\[
\langle \bar{q}q\rangle =-\frac{2N_{c}M_{q}}{\pi ^{4}{\rho _{c}}^{2}},\ \ \
\ \ \lambda _{q}^{2}=2\frac{1}{{\rho _{c}}^{2}}.
\]

By inverting these relations, we are able to express the parameters of the
instanton vacuum model in terms of the fundamental parameters of the QCD
vacuum
\begin{equation}
{\rho _{c}}^{2}=\frac{2}{\lambda _{q}^{2}},\ \ \ \ \ \
M_{q}=-\frac{\pi^{4}}
{N_{c}}\frac{\langle\bar{q}q\rangle}{\lambda _{q}^{2}} .
\label{19.1}
\end{equation}
>From these relations the model parameters, $\rho_{c}$ and $M_{q}$, can be
estimated in terms of the quark condensate,
$\langle \bar{q}q\rangle \approx-(230\ \mbox{MeV})^{3}$ \cite{Sch&Sh}
and the average quark virtuality. The
latter was estimated using QCD sum rules, $\lambda
_{q}^{2}=0.4\pm 0.2\ {\rm GeV}^{2}$ \cite{Bel&I}, and lattice QCD (LQCD)
calculations, which yield $\lambda _{q}^{2}=0.55\pm 0.05\ $GeV$^{2}$
\cite {Kra}.
These values lead to $\rho _{c}\approx 2$GeV$^{-1}$ and $M_{q}\approx
0.3$GeV. The combined analysis of the vacuum and pion properties confirms
such estimates \cite{DorL} and gives for the allowable range of parameters,
$1.5{\rm GeV}^{-1}\leq \rho _{c} \leq 2{\rm GeV}^{-1}$,
$0.22{\rm GeV} \leq M_{q} \leq
0.26$GeV. We note that the diluteness condition $\eta =M^{2}(\lambda
_{q}^{2})/\lambda _{q}^{2}\ll 1$ is well satisfied within the whole
{\sl window}.

\section{Deeply virtual Compton scattering and nonforward parton
distributions}

Let us consider the deeply virtual Compton scattering (DVCS) within our
model. The interest of such investigation is that it defines a procedure to
calculate the nonforward parton distributions, which correspond to a
generalization of the both, the parton distributions and the hadronic form
factors.

During virtual Compton scattering, a pion with momentum $p$ absorbs the
virtual photon of momentum $q$, producing an outgoing real photon of
momentum $q^{\prime }=q+r$ and a recoiling pion with momentum
$p^{\prime}=p-r$. We shall concentrate on the deeply virtual
kinematic region of $q$,
i.e., the Bjorken limit: $Q^{2}=-q^{2}\rightarrow \infty,\,p\cdot
q\rightarrow \infty$ and $Q^{2}/(p\cdot q)$, $t=r^2$ finite.

The expression for
the Compton scattering amplitude of the photon off the pion, with
momenta $p$, $q$ in the initial state and $p^{\prime}$, $q^{\prime}$
in the final state, is given by
\begin{eqnarray}
\label{4.4}
T_{\mu\nu}(p,q)&=&\int d\xi_1 d\xi_2 d\xi_3 d\xi_4 \exp\left(
iq\xi_1-iq^{\prime}\xi_2+ip\xi_3-ip^{\prime}\xi_4 \right)  \nonumber \\
&&\langle \frac{\delta^4 S}{\delta A_{\mu}(\xi_1)\delta A_{\mu}(\xi_2)
\delta M_{(\pi)}(\xi_3)\delta M_{(\pi)}(\xi_4)}\rangle_{0}.
\end{eqnarray}
The $S$-matrix has the standard form
\begin{eqnarray}  \label{5.4}
S=T\exp\left( i\int dx \left[ {\cal L}_{eff}(x)+{\cal L}_{em}(x)\right]
\right).
\end{eqnarray}
This amplitude corresponds to a set of diagrams drawn in figs.~$1$.
Fig. 1.1 shows the contribution of the $\sigma$ meson to the process
we are calculating, which is small, as can be shown numerically
and by $\frac{1}{N}$ counting\footnote{The direct $\sigma$ exchange diagram,
which would be leading order in 1/N, does not contribute to the isovector 
OFPD.}.
Moreover it can also
be shown that the
diagrams labelled ($c$) and ($d$) are suppressed in the Bjorken limit,
therefore we shall only pay attention to the diagrams ($a$) and ($b$).
Moreover, diagram ($a$) gives the biggest contribution. The general method
for obtaining of the off-diagonal distribution function can be found in
\cite{Rad},~\cite{Ji}. Following these methods,
we describe in detail only the contribution of diagram ($a$) to the
scattering amplitude in the momentum representation
\begin{eqnarray}
\label{13.4}
T_{\mu\nu}^{(\Delta)}(p,q)&=&\frac{g_{\pi\bar q q}^{2}}{4\pi^2} \int
\frac{d^{4}k}{4\pi^2i} f(k)f^2(k-p)f(k-r)\cdot
\nonumber \\
&&{\rm tr \biggl( \gamma_5 S(k-p)\gamma_5 S(k)\gamma_{\mu}
S(k+q)\gamma_{\nu} S(k-r) \biggr).}
\end{eqnarray}
In order to obtain the distribution functions,  it is convenient
to define a special coordinate system (see, for instance \cite{Ji}) in which
light-like vectors $p$ and $q^{\prime }$ are collinear. Then other vectors
are expanded in terms of $p$, $\hat{n}\equiv q^\prime/(p\cdot q^\prime)$ and
the transverse vectors
\footnote{
To define the parton distribution we formally turn to the Minkowski
signature.}:
\begin{eqnarray}  \label{14.4.1}
&&p^{\prime}_{\mu}=(1-\zeta) p_{\mu}-\frac{t}{2}\hat n_{\mu}+
p^{\prime\perp}_{\mu}, \quad r_{\mu}=\zeta p_{\mu}+\frac{t}{2}\hat
n_{\mu}+r^{\perp}_{\mu} ,  \nonumber \\
&&q_{\mu}=-\zeta p_{\mu}+(p\cdot q)\hat n_{\mu}+q^{\perp}_{\mu}, \quad
\zeta=r_{\parallel}\cdot\hat n=1-p^{\prime}\cdot\hat n,  \nonumber \\
&&p\cdot q=p\cdot q^{\prime}+\frac{t}{2},
\end{eqnarray}
where $\zeta $ is the asymmetry parameter which in the Bjorken limit
coincides with the Bjorken variable $x_{B}$ for deep-inelastic scattering
\[
x_{B}=-\frac{q^2}{2pq}=\frac{Q^2}{2pq}.
\]

Now we insert the given decompositon of $4$-vectors in eq. (\ref{13.4}) and
carry out the integration over $k$, keeping only
terms which are not suppressed in the Bjorken limit.
Then, using
\begin{eqnarray} \label{16.4} \int\limits_{-1}^{1} d\tilde X
\delta(\tilde X-k\hat n)=1,
\end{eqnarray}
we obtain for the contribution of diagram ($a$) th the
Compton scattering amplitude
\begin{eqnarray}
\label{17.4}
T_{\mu \nu }^{\Delta }(q,q^{\prime },r)=
\frac{1}{2}\left( p_{\mu }\hat{n}_{\nu }+
\hat{n}_{\mu }p_{\nu }-g_{\mu \nu }\right) \int dX\frac{1}{X-\zeta }
{\cal F} _{\zeta }^{\Delta }(X,t)
\end{eqnarray}
where

\begin{eqnarray}
&&{\cal F}_{\zeta }^{\Delta }(X,t)=\frac{1}{2-\zeta }\frac{g_{\pi
\bar{q}
q}^{2}}{4\pi ^{2}}\int \frac{d^4 k}{2(4\pi ^{2}i)}f(k)f^{2}(k-p)f(k-r)
\nonumber  \label{26} \\
&&{\rm tr}\biggl(\gamma _{5}S(k-p)\gamma _{5}S(k)\hat{n}\cdot
\gamma S(k-r) \biggr)
\delta (X-k\hat{n})+(X\rightarrow \bar{X}).
\end{eqnarray}

In a similar manner we obtain the contribution of diagram ($b$) which
is given by
\begin{eqnarray}
\label{17.4.1}
T_{\mu \nu }^{\circ }(q,q^{\prime },r)=
\frac{1}{2}\left( p_{\mu }\hat{n}_{\nu }+
\hat{n}_{\mu }p_{\nu }-g_{\mu \nu }\right) \int dX\frac{1}{X-\zeta }
{\cal F}_{\zeta }^{\circ } (X,t),
\end{eqnarray}
where
\begin{eqnarray}
{\cal F}_{\zeta }^{\circ }(X,t) &=&\frac{1}{2-\zeta }\frac{g_{\pi
\bar{q}
q}^{2}}{4\pi ^{2}}\int \frac{d^4 k}{4\pi ^{2}i}f(k)f^{2}(k+p^{\prime
})\int\limits_{0}^{1}d\tau f^{\prime }((k-r\tau )^{2}+r^{2}\tau
(1-\tau ))
\nonumber  \label{27} \\
& &{\rm tr}\Biggl(\gamma _{5}S(k)\gamma _{5}S(k+p^{\prime })\Biggr)
\Biggl(2X-\frac{3}{2}\zeta \Biggr)
\nonumber \\
&&\Biggl(\delta (X-k\hat{n})+\delta (X-\zeta +k\hat{n})\Biggr)+
(X\rightarrow\bar{X}) .
\end{eqnarray}

The variable $X$ is the total fraction of the initial hadron momentum $p$
carried by the active quark, and satisfies the kinematical constraint
$0\leq X\leq 1$.

In order to calculate the structure integrals in eqs. (\ref{26}) and
(\ref{27}) we use the $\alpha-$representation for all propagators,
the Laplace
transform for the nonlocal vertex function $f(k)$ and the integral
representation for $\delta-$function. Note that, in the $k-$integrals which
appear in the above equations, we use the Euclidean signature, but at the
final stage we move back to Minkowski space for the external momenta. In the
case of $t=0$ we consider a family of asymmetric
distribution functions ${\cal F}_{\zeta }(X)$ whose shapes change
with $\zeta $. The $\zeta $
-dependence of the asymmetric function is the main difference with respect
to the double distribution function $F(x,y)$ recently introduced in
~\cite{Rad}. The asymmetric function ${\cal F}_{\zeta }(X)$
satisfies the sum-rules,
\begin{equation}
\int\limits_{0}^{1}dX{\cal F}_{\zeta }(X)=1.
\end{equation}
This distribution function has the following properties
\cite{Rad}: When the total momentum fraction $X$ of the initial hadron
momentum $p$ is larger than the fraction $\zeta $ of the momentum transfer
$r$, the function ${\cal F}_{\zeta }(X)$ can be treated as a
generalization of the usual distribution function;
when $X<\zeta $, the asymmetric function can be
treated like a distribution amplitude $\Phi $ of the $\bar{q}q$-states, in
our case the pion, with momentum $\zeta p$;
besides, for $\zeta =1$, the asymmetric quark
distribution function ${\cal F}_{\zeta }(X) $ becomes the pion wave function
(see, for example, \cite{Pol}).

The asymmetric quark distribution functions ${\cal F}_{\zeta}^{\Delta}$ and $%
{\cal F}_{\zeta}^{\circ}$ are shown in fig.~2.

Finally, the integration of Eqs. (\ref{26}), (\ref{27}) over the total
momentum fraction $X$ yields the following sum rule:
\begin{equation}  \label{28}
\int\limits_{0}^{1}dX{\cal F}_{\zeta }(X,t)=F_{\pi }(t),
\end{equation}
where $F_{\pi }(t)$ is the hadronic form factor normalized to unity. The
behavior of the form factor $F_{\pi }(t)$ is shown in fig.~3. Our numerical
results are in agreement with experimental data and also coincide with
results given by other approaches (see, for example, \cite{Ani}).

\section{Concluding remarks}

We have used the instanton liquid model of the QCD vacuum to obtain an
effective quark-hadron lagrangian by bosonizing the 4-fermion interaction
generated by the instantons. The procedure imposes a coupling constant
condition which allows us to calculate the physical value of the
quark-hadron interaction constant.

The parameters of the model, related to the properties of the vacuum
structure, have been fixed by using the $<\bar{q} q >$ condensate and the
average quark virtuality of the vacuum. Furthermore, the pion properties
confirm our estimates.

We have considered within the model deeply virtual Compton scattering a
procedure which allows the calculation of nonforward parton distributions.
We have calculated the asymmetric distribution functions at different
asymmetrical parameters ($\zeta$).
Finally the
integration over the total momentum fraction on the nonforward parton
distribution yields the pion form factor. Our results agree with those of
other authors and with experiments \cite{Beb}.

This work represents step forward in the direction of deriving a model
for hadron structure directly from $QCD$ with well established
assumptions. The procedure developed can be generalized to other mesons
and to baryons, and the work in this direction is in progress.

\section*{Acknowledgments}
\vspace{2mm}
A.E.D. thanks his colleagues from Instituto
de F\'{\i}sica Te\'{o}rica, UNESP, (S\~{a}o Paulo) for their
hospitality and interest in this work.
I.V.A. thanks colleagues from Departamento de Fisica
Teorica of Universidad de Valencia
for warm hospitality and very useful discussions.
This investigation
(I.V.A., A.E.D. and A.E.M.) was supported in part by the
Russian Foundation for Fundamental Research (RFFR) 96-02-18096
and 96-02-18097, St. - Petersburg Center for Fundamental
Research grant: 97-0-6.2-28. L.T. also thanks partial support
received from Funda\c{c}\~{a}o de Amparo \`{a} Pesquisa do
Estado de S\~{a}o Paulo (FAPESP) and from Conselho Nacional
de Desenvolvimento Cient\'{\i}fico e Tecnol\'{o}gico do
Brasil. I.V.A. was supported during his visit to Valencia by el Acuerdo
de Intercambio Universidad de Valencia-JINR y DGICYT
grant PB97-1227.

\newpage
\vspace*{0.5cm}
\begin{center}
{\bf Appendix $A$}
\end{center}
\vspace*{0.5cm}
\indent

\noindent 1. The contribution arising from the basic diagrams (see, fig.~$1$a):
\begin{eqnarray}
{\cal F}_{\zeta}^{\Delta}(X,t)=\frac{1}{2-\zeta}
\frac{g_{\pi\bar q q}^{2}}{4\pi^2}
{\cal I}_{\zeta}^{\Delta}(\alpha,t)
\nonumber
\end{eqnarray}
where
\begin{eqnarray}
{\cal I}_{\zeta}^{\Delta}(\alpha,t)&=&
\frac{1}{2}
\int\limits_{0}^{\infty} d\alpha_1..d\alpha_4
\tilde f_1(\alpha_1)\tilde f_2(\alpha_2)\tilde f_1(\alpha_3)
e^{\alpha_4 M^2}
\nonumber\\
&&\Biggl(
\frac{X e^{\tilde\alpha_1 M^2}\Theta (\tilde\alpha_1)}
{\alpha_2+\alpha_{34}\zeta}
{\rm exp}\left( -tX\frac{\alpha_{34}(\alpha_1+\tilde\alpha_1)}
{\alpha_2+\alpha_{34}\zeta}\right)+
\Biggr.
\nonumber\\
&&\Biggl.
\frac{(1-\zeta) e^{\tilde\alpha_2 M^2}\Theta (\tilde\alpha_2)}
{\alpha_1+\alpha_{34}(1-\zeta)}
{\rm exp}\left( -t(1-X)\frac{\alpha_{34}\alpha_1}
{\alpha_1+\alpha_{34}(1-\zeta)}\right)
\Biggr) -
\nonumber\\
&&\frac{1}{2}\int\limits_{0}^{\infty} d\alpha_1..d\alpha_5
\tilde f_1(\alpha_1)\tilde f_2(\alpha_2)\tilde f_1(\alpha_3)
\frac{(1-\zeta)(1-X)t\alpha_{35} \Theta (\tilde\alpha_3)}
{\left( \alpha_{14}+\alpha_{35}(1-\zeta) \right)^2}
\nonumber\\
&&{\rm exp}\left( -M^2(\alpha_{45}+\tilde\alpha_3)
-t(1-X)\frac{\alpha_{35}\alpha_{14}}
{\alpha_{14}+\alpha_{35}(1-\zeta)} \right).
\nonumber
\end{eqnarray}

\noindent 2. The contribution arising from the tadpole diagrams (see,
fig.~$1$b):
\begin{eqnarray}
{\cal F}_{\zeta}^{\circ}(X,t)=\frac{1}{2-\zeta}
\frac{g_{\pi\bar q q}^{2}}{4\pi^2}
\left( {\cal I}_{\zeta}^{\circ (1)}(\alpha,t)+
{\cal I}_{\zeta}^{\circ (2)}(\alpha,t) \right) ,
\nonumber
\end{eqnarray}
where
\begin{eqnarray}
{\cal I}_{\zeta}^{\circ (1)}(\alpha,t)&=&
\int\limits_{0}^{\infty} d\alpha_1..d\alpha_3
\alpha_3\tilde f_1(\alpha_1)\tilde f_2(\alpha_2)\tilde f_1(\alpha_3)
\int\limits_{0}^{1} d\tau
\frac{2X-1.5\zeta}{\alpha_3\zeta\tau-\alpha_2(1-\zeta)}
\nonumber\\
&&\Biggl(
\Theta(\hat\alpha_1){\rm exp}\left(
\hat\alpha_1 M^2-t\tau\alpha_3
\frac{\alpha_1+\hat\alpha_1+\alpha_3(1-\tau)}
{\alpha_{123}+\hat\alpha_1} \right)+
\Biggr.
\nonumber\\
\Biggl.
&&\Theta(\hat\alpha_2){\rm exp}\left(
\hat\alpha_2 M^2-t\tau\alpha_3
\frac{\alpha_1+\hat\alpha_2+\alpha_3(1-\tau)}
{\alpha_{123}+\hat\alpha_2} \right)
\Biggr),
\nonumber
\end{eqnarray}

\begin{eqnarray}
{\cal I}_{\zeta}^{\circ (2)}(\alpha,t)&=&
\int\limits_{0}^{\infty} d\alpha_1..d\alpha_3
\alpha_3\tilde f_1(\alpha_1)\tilde f_2(\alpha_2)\tilde f_1(\alpha_3)
\int\limits_{0}^{1} d\tau
\frac{2X-1.5\zeta}{\alpha_{13}(1-\zeta)+\alpha_3\zeta\tau}
\nonumber\\
&&\Biggl(
\Theta(\hat\alpha_3){\rm exp}\left(
\hat\alpha_3 M^2-t\tau\alpha_3
\frac{\alpha_1+\alpha_3(1-\tau)}
{\alpha_{123}+\hat\alpha_3} \right)+
\Biggr.
\nonumber\\
\Biggl.
&&\Theta(\hat\alpha_4){\rm exp}\left(
\hat\alpha_4 M^2-t\tau\alpha_3
\frac{\alpha_1+\alpha_3(1-\tau)}
{\alpha_{123}+\hat\alpha_4} \right)
\Biggr).
\nonumber
\end{eqnarray}
The notation is:
$$
\alpha_{n_1..n_i}=\alpha_{n_1}+..+\alpha_{n_i} ,
$$
$$
\tilde\alpha_1=\alpha_2\frac{1-X}{X}-\alpha_{34}\frac{X-\zeta}{X}-\alpha_1
,
$$
$$
\tilde\alpha_2=\alpha_1\frac{X}{1-X}+\alpha_{34}\frac{X-\zeta}{1-X}-\alpha_2,
$$
$$
\tilde\alpha_3=\alpha_{14}\frac{X}{1-X}+\alpha_{35}\frac{X-\zeta}{1-X}
-\alpha_2
,
$$
$$
\hat\alpha_1=\alpha_3\frac{\zeta\tau-X}{X}-
\alpha_2\frac{1-\zeta+X}{X}-\alpha_1
,
$$
$$
\hat\alpha_2=-\alpha_3\frac{\zeta(1-\tau)-X}{\zeta-X}-
\alpha_2\frac{1-X}{\zeta-X}-\alpha_1
,
$$
$$
\hat\alpha_3=-\alpha_3\frac{\zeta(1-\tau)-X}{1-X}-
\alpha_1\frac{\zeta-X}{1-X}-\alpha_2
,
$$
$$
\hat\alpha_4=-\alpha_3\frac{X-\zeta\tau}{1-\zeta+X}-
\alpha_1\frac{X}{1-\zeta+X}-\alpha_2.
$$

\newpage

\begin{figure}
\begin{center}
\includegraphics[width=8cm]{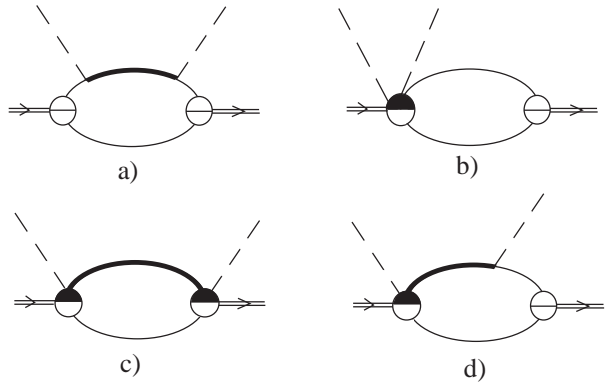}
\caption{The diagrams contributing to the Compton Scattering amplitude:
$(a)$ is the basic type and $(b)-(d)$ are tadpoles.}
\end{center}
\end{figure}

\newpage

\begin{figure}
\begin{center}
\includegraphics[width=8cm]{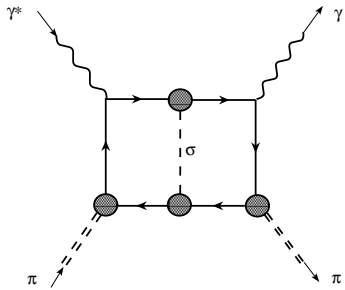}
\caption{The sigma exchange diagrams contributing to the
Compton Scattering amplitude.}
\end{center}
\end{figure}

\newpage

\begin{figure}
\begin{center}
\includegraphics[width=10cm]{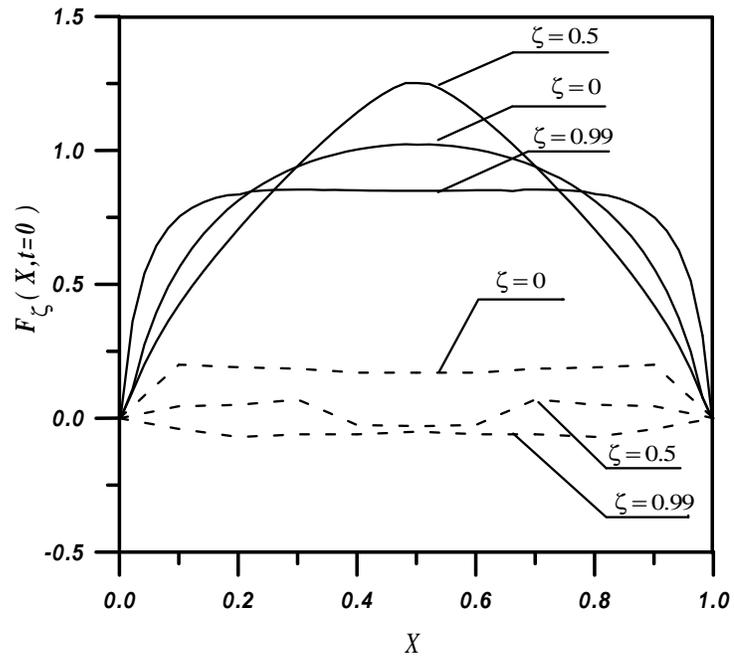}
\caption{The asymmetric distribution functions at different values of
an asymmetry parameter $\zeta$.
The solid line shows the contribution
to the distribution function of both the basic and the tadpole diagrams;
the dashed-line represents the contribution from the tadpoles alone.}
\end{center}
\end{figure}

\newpage

\begin{figure}
\begin{center}
\includegraphics[width=10.0cm]{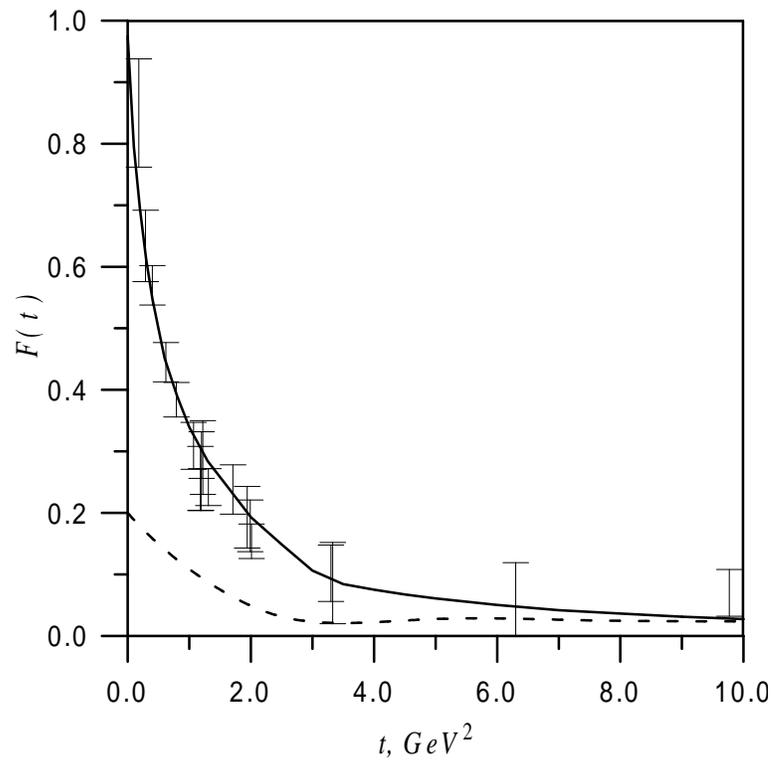}
\caption{The pion elctromagnetic form factor. The solid line shows
the contribution from both the basic and tadpole diagrams, the dashed-line
again represents the contribution from the tadpoles only.}
\end{center}
\end{figure}

\end{document}